\documentclass[twoside,epsf]{article}
\usepackage{amsmath,amsfonts}
\usepackage{qic,epsfig}
\usepackage{cite}

\newcommand{\RE}{\mbox{Re}}
\newcommand{\IM}{\mbox{Im}}

\renewenvironment{proof}{\par\noindent{\bf Proof.}}{$\quad\Box$\par}

\begin{document}
\setlength{\textheight}{8.0truein}    

\runninghead{A new algorithm for producing quantum circuits using
$KAK$ decompositions}
            {Yumi Nakajima, Yasuhito Kawano, and Hiroshi Sekigawa}

\normalsize\textlineskip
\thispagestyle{empty}
\setcounter{page}{1}

\copyrightheading{0}{0}{2003}{000--000}

\vspace*{0.88truein}

\alphfootnote

\fpage{1}

\centerline{\bf A new algorithm for producing quantum circuits}
\vspace*{0.035truein}
\centerline{\bf using $KAK$ decompositions}
\vspace*{0.37truein}
\centerline{\footnotesize 
Yumi Nakajima\footnote{nakajima@theory.brl.ntt.co.jp},
Yasuhito Kawano, and Hiroshi Sekigawa}
\vspace*{0.015truein}
\centerline{\footnotesize\it 
NTT Communication Science Laboratories, 
Nippon Telegraph and Telephone Corporation}
\baselineskip=10pt
\centerline{\footnotesize\it 3-1 
Morinosato-Wakamiya, Atsugi, Kanagawa 243-0198
, JAPAN}
\vspace*{0.225truein}
\publisher{(received date)}{(revised date)}

\vspace*{0.21truein}
\abstracts{
We provide a new algorithm that translates a unitary matrix into a
quantum circuit according to the $G=KAK$ theorem in Lie group theory.
With our algorithm, any matrix decomposition corresponding to
type-{\bf AIII} $KAK$ decompositions can be derived according to the
given Cartan involution.
Our algorithm contains, as its special cases, Cosine-Sine
decomposition (CSD) and Khaneja-Glaser decomposition (KGD)
in the sense that it derives the same quantum circuits
as the ones obtained by them if we select suitable Cartan involutions
and square root matrices.
The selections of Cartan involutions for computing CSD and KGD 
will be shown explicitly. 
As an example, we show explicitly that our method can automatically 
reproduce the well-known efficient quantum circuit for the $n$-qubit 
quantum Fourier transform.   
}{}{}

\vspace*{10pt}
\keywords{$G=KAK$ theorem, KGD, CSD, decomposition of the QFT}
\vspace*{3pt}
\communicate{to be filled by the Editorial}

\vspace*{1pt}\textlineskip	

\section{Introduction}\label{sec:intro}
\noindent 
Decomposing a unitary matrix into an efficient sequence of elementary
gates  is a fundamental problem in  designing quantum circuits. 
There are two types of decomposition: One is exact decomposition where
an arbitrary unitary matrix is decomposed precisely into a sequence of
elementary gates, such as arbitrary single-qubit rotations and
the CNOT. The other involves approximate strategies by which an arbitrary
unitary matrix is decomposed approximately into a sequence of a fixed
set of elementary gates, as shown in Solovay-Kitaev
theorem~(cf.~\cite{NC}, Appendix~3).  
In this paper, we treat exact decomposition.

Cosine-Sine decomposition (CSD), which is a well-known algorithm in
numerical linear algebra, was the first algorithm utilized 
for this purpose~\cite{tucci1}.
CSD applies the well-known algorithm for computing generalized
singular value  decomposition (GSVD).
In CSD, we first divide an input matrix $g$ into four square matrices 
and then apply SVD to each matrix.
Then, we have
\begin{eqnarray}
g=\begin{pmatrix} 
g_{11} & g_{12} \\
g_{21} & g_{22} 
\end{pmatrix}
& = & \begin{pmatrix}
u_1 & 0 \\
0 & u_2 
\end{pmatrix}
\begin{pmatrix}
c & -s \\
s & c 
\end{pmatrix}
\begin{pmatrix}
v_1 & 0 \\
0 & v_2 
\end{pmatrix}\: = \: U\Sigma V, \label{def-csd}
\end{eqnarray}
where $c = diag \left[ \cos(\zeta_1),\cos(\zeta_2),\cdots,
\cos(\zeta_{2^{n-1}}) \right]$,
$s = diag \left[ \sin(\zeta_1),\sin(\zeta_2),\cdots
  ,\sin(\zeta_{2^{n-1}}) \right]$, then $g_{11} = u_1 c v_1$,
$g_{12}=-u_1 s v_2$, $g_{21}=u_2 s v_1$, and $g_{22}=u_2 c v_2$ are
SVD.  
Recursively performing decomposition (\ref{def-csd})
leads to a quantum circuit.
The CSD-based algorithms are easy to implement on a computer
because algorithms for calculating GSVD are well-known, 
and software libraries including GSVD are available.
Some CSD-based algorithms~\cite{svore2, hut1, hut2, shende2}
have been investigated with the aim of improving Barenco's result
that an arbitrary $2^n \times 2^n$ unitary matrix is composed of
$O(n^24^n)$ elementary gates~\cite{barenco}. 
And improvement to $O(4^n)$ elementary gates has been reported by
M\"{o}tt\"{o}nen \textit{et al}.~\cite{hut1} and by Shende \textit{et
  al}~\cite{shende2}.  

On the other hand, Khaneja and Glaser provided another kind of
decomposition~\cite{khaneja1}, which was later named KGD.
KGD lies within the framework of the $G=KAK$ theorem 
(cf.~\cite{helgason}, Theorem~8.6) in Lie group theory. 
This theorem shows that an element $g\in SU(2^n)$ is decomposed 
into matrix products $k_1ak_2$ for some 
$k_1, k_2 \in \exp(\mathfrak{k})$ and $a \in \exp(\mathfrak{h})$. 
Here, $\mathfrak{su}(2^n)=\mathfrak{k} \oplus \mathfrak{m}$ is 
a Cartan decomposition in Lie algebra $\mathfrak{su}(2^n)$, 
$\mathfrak{k}$ and $\mathfrak{m} =\mathfrak{k}^\bot$ are 
orthogonal vector spaces contained in $\mathfrak{su}(2^n)$, 
and $\mathfrak{h}$ is a maximal Abelian subalgebra 
(a Cartan subalgebra) contained in $\mathfrak{m}$ 
(cf.~\cite{knapp}, \S VI.2).
Matrices $k_1$, $a$, and $k_2$ are not uniquely determined from $g$.
They depend on the selections of the bases of $\mathfrak{k}$,
$\mathfrak{m}$, and $\mathfrak{h}$; besides, they are not determined
even if bases are selected.
Khaneja and Glaser provided a particular selection of bases of
$\mathfrak{k}$, $\mathfrak{m}$, and $\mathfrak{h}$ in
Ref.~\cite{khaneja1} so that the selection matches an NMR system, 
and they proved that a time-optimal control on a two-qubit NMR quantum
computer can be obtained from the decomposition~\cite{khaneja2}. 
Thus, KGD can be regarded as the $G=KAK$ theorem on the 
particular bases. It should be noted that KGD does not give a 
unique translation of the input matrix into a quantum circuit.

Bullock~\cite{bullock1} showed that CSD can also be regarded
in the framework of the $G=KAK$ theorem; i.e.,
CSD uses the type-{\bf AIII} $KAK$ decomposition with the global 
Cartan decomposition $\Theta$ defined as 
$\Theta(X)=\sigma_{1z}X\sigma_{1z}$ for $X \in SU(2^n)$, 
where $\sigma_{jz}$ denotes that the operation defined as the 
Pauli matrix $\sigma_{z}$ acts on the $j$-th qubit. 
He also introduced a method that translates matrices $U$, $\Sigma$,
and $V$ in (\ref{def-csd}) into 
$k_1 \in \exp(\mathfrak{k})$, $a \in \exp(\mathfrak{h})$, and
$k_2 \in \exp(\mathfrak{k})$, respectively, 
where $\mathfrak{k}$ and $\mathfrak{h}$ are the ones defined in KGD. 
Here, KGD corresponds to $G=KAK$ decomposition with the 
selection of $\Theta$ defined as $\Theta(X)=\sigma_{nz}X\sigma_{nz}$. 
We can thus produce a KGD-based quantum circuit
by combining Bullock's translation and the CSD-based algorithms.

We introduce a new algorithm that translates a $2^n \times 2^n$
unitary matrix into a quantum circuit according to the $G=KAK$ theorem.
The algorithm can derive any matrix decomposition corresponding to 
type-{\bf AIII} $KAK$ decompositions for the given global Cartan 
involution $\Theta$. 
The algorithm contains, as its special cases, both CSD and
KGD in the sense that it derives the same quantum
circuits as the ones calculated by them
if we select suitable Cartan involutions and square root matrices. 
Here, we select $\Theta(X)$ as $\sigma_{1z} X \sigma_{1z}$ for CSD and 
as $\sigma_{nz} X \sigma_{nz}$ for KGD, where $X \in SU(2^n)$. 
The strategy utilized in our algorithm is related to those used in
Refs.~\cite{bullock2, bullock3}. However, those strategies provided 
methods for computing type-{\bf AII} $KAK$ decomposition; 
no translation between type-{\bf AII} decompositions and 
type-{\bf AIII} decompositions was provided. 
Furthermore, the method utilized in Ref.~\cite{bullock2} is different
from ours in the square root matrix calculations, i.e., methods for 
calculating $m$ from $m^2$ (where $g=km$ is a global Cartan 
decomposition of the input matrix $g$). 
In the method proposed in Ref.~\cite{bullock2}, first, 
a square root matrix is calculated in Lie algebra level. 
And then it is translated into an element in Lie group level via 
exponential mapping. In contrast, with our method, a square root 
matrix is calculated directly at the Lie group level. 

Although our algorithm contains CSD and derives any matrix 
decomposition corresponding to type-{\bf AIII} 
$KAK$ decompositions according to the given Cartan involution, the
efficiency for calculating a decomposition is not sacrificed.
The reason is as follows: 
Roughly speaking, to decompose $g$ into $k1ak2$, the CSD-based 
algorithms apply SVD to four $2^{n-1}\times 2^{n-1}$ matrices 
($g_{11}$, $g_{12}$, $g_{21}$, and $g_{22}$ in Eq.(\ref{def-csd})),
while our algorithm applies eigenvalue decomposition to 
$2^{n} \times 2^{n}$ matrix. 
Therefore, the efficiencies for computing SVD on four 
$2^{n-1} \times 2^{n-1}$ matrices and for computing eigenvalue 
decomposition on one $2^{n} \times 2^{n}$ matrix are the same.

In addition, our algorithm might have an advantage over CSD
when we would like to determine a class of quantum circuits
for a given class of matrices. The reason is as follows: 
In CSD-based algorithms, it is difficult to formulate a class of
matrices $u_1$, $u_2$, $v_1$, and $v_2$ such that relation
(\ref{def-csd}) holds for a given class of input matrices. 
Actually, to reproduce the well-known QFT circuit by using
CSD~\cite{tucci1, tucci2}, 
Tucci changes the rows and columns of the QFT matrices beforehand 
and makes each submatrix hold a convenient form, which can be written
by the $(n-1)$-qubit QFT. 
It would not be possible to describe the general form of the
decomposition when the input matrix does not have a convenient form
like QFT. 
On the other hand, our algorithm does not require such a preliminary change
of rows and columns. All matrices appearing through our algorithm can
be described using the input matrix $g$, the given global Cartan involution
$\Theta$, and the eigenvalues and eigenvectors of these matrix
products. This will be shown explicitly as an example of the QFT 
decomposition (Section~4).

The paper is organized as follows: In the following section, we cover
 some preliminaries about notations, the $G=KAK$ theorem, and
 KGD. Section~\ref{sec:kak-alg} presents our algorithm for computing
 the decomposition follows from the $G=KAK$ theorem. Section~\ref{sec:QFT}
 presents decompositions of the $n$-qubit QFT using our algorithm
 and CSD-based algorithms. We show that we can produce the well-known QFT
 circuit by using these matrix decompositions. 

\section{Preliminaries}\label{sec:prelim}
\subsection{Notations}\label{sec:notation}
\noindent
Let $\sigma_x$, $\sigma_y$, and $\sigma_z$ denote the Pauli matrices  
and $I^{\otimes s}$ be a $2^s \times 2^s$ identity matrix
($I=2^1 \times 2^1$). 
We use $\sigma_{j\alpha}$ to denote the Pauli matrix acting on the
$j$-th qubit; $\sigma_{j \alpha}=I^{\otimes (j-1)}\otimes
\sigma_{\alpha} \otimes I^{\otimes (n-j)}$, ($\alpha=x,y$, or
$z$). Let $U_{\mathrm{CNOT}}$ denote the standard CNOT gate, $H$
denote a Hadamard gate, and $R_x(\zeta)=\exp (-i \zeta \sigma_x)$.
All these notations follow those in Ref.~\cite{NC}.

\subsection{$G=KAK$ theorem} \label{sec:kak-theorem}
\noindent
The $G=KAK$ theorem for compact groups
(cf.~\cite{helgason}, Theorem~8.6)
provides a framework for decomposing $g \in SU(2^n)$ into the following
matrix products:   
\begin{eqnarray}
g & = & k_1 a k_2, \qquad k_1, k_2 \in \exp(\mathfrak{k}), a \in
\exp(\mathfrak{h}) \subset \exp(\mathfrak{m}). \label{kak}
\end{eqnarray}
Here, $\mathfrak{su}(2^n)=\mathfrak{k} \oplus \mathfrak{m}$ is a
Cartan decomposition, where $\mathfrak{k}$ and
$\mathfrak{m}=\mathfrak{k}^\bot$ are orthogonal vector spaces, 
and $\mathfrak{h}$ is a Cartan subalgebra, that is, a maximal Abelian
subalgebra contained in $\mathfrak{m}$.  

Let $\theta$ denote the Cartan involution of its Lie algebra
$\mathfrak{su}(2^n)$; i.e.,  
(i) $\theta^2 = I^{\otimes n}$ ($\theta
\neq I^{\otimes n}$) and 
(ii) $\theta$ is an automorphism of the Lie algebra
$\mathfrak{su}(2^n)$. And let the global Cartan 
involution (cf.~\cite{knapp}, p.~362) of $SU(2^n)$ be $\Theta$. Then  
$\mathfrak{k}$ and $\mathfrak{m}$ have the following property: 
\begin{alignat}{2}
\theta(x) &= \begin{cases}
x & \quad\mbox{if $x \in \mathfrak{k}$}\\
-x & \quad\mbox{if $x \in \mathfrak{m}$}\\
\end{cases},
& \qquad
\Theta(X) &= \begin{cases}
X & \quad\mbox{if $X \in \exp(\mathfrak{k})$}\\
X^\dag & \quad\mbox{if $X \in \exp(\mathfrak{m})$}\\
\end{cases}. \label{property1}
\end{alignat}
Three types of $\mathfrak{k}$-algebra, named {\bf AI}, 
{\bf AII}, and {\bf AIII}, arise for $\mathfrak{su}(2^n)$. 
Here, {\bf AI}, {\bf AII}, and {\bf AIII} correspond to  
$\mathfrak{k}=\mathfrak{so}(2^n)$,
$\mathfrak{k}=\mathfrak{sp}(2^n)$, and
$\mathfrak{s}[\mathfrak{u}(p) \oplus \mathfrak{u}(q)]$ ($p+q=2^n$), 
respectively~(cf.~\cite{helgason}, p.~518).

\subsection{Khaneja-Glaser decomposition (KGD)} \label{sec:kgd}
Khaneja and Glaser provided a particular selection of bases of
$\mathfrak{k}$, $\mathfrak{m}$, and
$\mathfrak{h}$~(cf.~\cite{khaneja1}, Notation~3 and 5) 
so that the selection matches an NMR system. Notice that we
use $\mathfrak{h}$ instead of $\mathfrak{h}(n)$.
Here, generators of $\mathfrak{k}$, $\mathfrak{m}$, and
$\mathfrak{h}$, are denoted to be tensor products of the Pauli
matrices;
\begin{eqnarray}
\mathfrak{k} & = & \mathrm{span}\: \{A\otimes \sigma_z/2, B\otimes
I, i \sigma_{nz}/2 \: | \: A, B \in \mathfrak{su}(2^{n-1})
\}, \label{defK}\\
\mathfrak{m} & = & \mathrm{span}\: \{A\otimes \sigma_x/2, B\otimes
\sigma_{y}/2, i \sigma_{nx}/2, i \sigma_{ny}/2\: | \: A, B \in
\mathfrak{su}(2^{n-1}) \}. \label{defM}
\end{eqnarray}
Here, generators of $\mathfrak{k}$ and $\mathfrak{m}$ have a specific
operation on the last qubit; i.e., $\sigma_{z}$ or $I$ for
generators of $\mathfrak{k}$ and $\sigma_{x}$ or $\sigma_{y}$ for
generators of $\mathfrak{m}$. Thus, to determine $\mathfrak{k}$ and
$\mathfrak{m}$, the Cartan involution $\theta$ and the global Cartan
involution $\Theta$ can be chosen as follows: 
\begin{alignat}{2}
\theta(x) &= \sigma_{nz}x \sigma_{nz}, &\qquad \qquad
\Theta(X) &= \sigma_{nz} X \sigma_{nz}. \label{def-theta}
\end{alignat}
Since $\theta(\sigma_z)=\sigma_z$, $\theta(I)=I$,
$\theta(\sigma_x)=-\sigma_x$, $\theta(\sigma_y)=-\sigma_y$, we can
check that the above $\mathfrak{k}$ and $\mathfrak{m}$ satisfy 
relation (\ref{property1}) when $\theta$ and $\Theta$ are chosen as
(\ref{def-theta}). 
 
\begin{figure}[tbp]
\vspace*{10pt}
\centerline{\epsfig{file=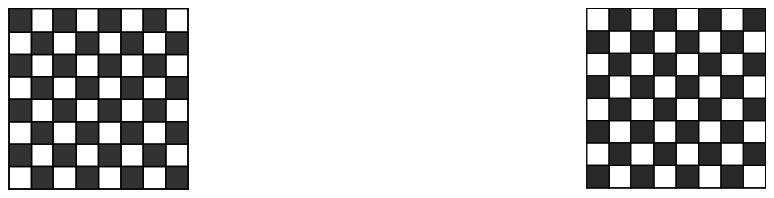}}
\centerline{\hspace{-0.5cm} $\mathfrak{k}$ and $\exp(\mathfrak{k})$
  \hspace{4.5cm} $\mathfrak{m}$}
\vspace{0.3cm}
\fcaption{Patterns of an element of $\mathfrak{k}$, $\exp(\mathfrak{k})$ 
and $\mathfrak{m}$ for a three-qubit system, where each square
represents an element of an $8\times 8$ matrix. The white elements are
always zero, and the black elements take some value that depends on
the input matrix. \label{pattern}} 
\end{figure}

For the number of qubits $n \geq 3$, $\mathfrak{k}$ and $\mathfrak{m}$
have specific patterns, as shown in Fig.~\ref{pattern}, because all
generators defined in (\ref{defK}) and (\ref{defM}) have these
patterns. Note that in contrast to an element of $\exp(\mathfrak{k})$
taking the same pattern as an element of $\mathfrak{k}$, an element of
$\exp(\mathfrak{m})$ does not take the same pattern as $\mathfrak{m}$.  
This property enables us to apply the $KAK$ decomposition recursively,
as shown in Fig.~\ref{msd-image}. 

\section{Our algorithm \label{sec:kak-alg}}
\subsection{Basic ideas}\label{sec:alg} 
\noindent
We provide a new constructive algorithm that computes a decomposition
based on the $G=KAK$ theorem. Here, we choose a Cartan subalgebra
$\widetilde{\mathfrak{h}}$, which is different from the $\mathfrak{h}$
used in KGD.  
Since Cartan subalgebras are Abelian, they can
translate each other by  
$\widetilde{\mathfrak{h}}=Ad_{T \in \exp(\mathfrak{k})}
(\mathfrak{h})$. Here, $T$ is fixed for given
$\widetilde{\mathfrak{h}}$ and $\mathfrak{h}$.
It should be noted that $G=KAK$ decomposition for the fixed input $g \in
SU(2^n)$ is not unique. We provide the following theorem. 

\vspace*{12pt}
\begin{theorem}
Let $g \in SU(2^n)$ be the input matrix. 
If $g$ has a global Cartan decomposition $g=km \;(k \in \exp(\mathfrak{k}), 
m \in \exp(\mathfrak{m}))$, then $m^2$ is uniquely determined by 
$m^2 = \Theta(g^\dag) g$.
\end{theorem} 

\vspace*{12pt}
\noindent
\begin{proof}
From (\ref{property1}), 
  $\Theta(g^\dag)g = \Theta(m^\dag k^\dag) km
 = \Theta(m^\dag)\: \Theta(k^\dag)km = mk^\dag km = m^2$. 
\end{proof} 
\vspace*{12pt}
\noindent
 
Theorem~1 shows that the fixed global Cartan involution $\Theta$ only
determines $m^2$. Therefore, arbitrariness remains in the selection of $m$,
and also $k$. Furthermore, $a$ in (\ref{kak}) has also arbitrariness
because it follows from a decomposition of $m$, 
$m = \widetilde{k}^\dag a \widetilde{k}$, 
where $\widetilde{k} \in \exp(\mathfrak{k})$.
(See, \cite{knapp}, \S AII.3). 
Using Theorem~1, we compute the decomposition in (\ref{kak}) as
follows: 
 
\begin{description}
\item{1.} Compute $m^2=\Theta(g^\dag)g$. 
\item{2.} Decompose $m^2=pbp^\dag$ such that $p \in
  \exp(\mathfrak{k})$ and $b \in \exp(\widetilde{\mathfrak{h}})$. \\
Such decomposition always exists because $m^2 \in
  \exp(\mathfrak{m})$ (cf.~\cite{knapp}, Proposition~7.29). We compute
  $p$ using eigenvectors of $m^2$. We show examples in
  Section~\ref{sec:detail}. 
\item{3.} Find $y$ such that $y^2=b$ and $y\in
  \exp(\widetilde{\mathfrak{h}})$. \\
The $y$ can be computed by replacing the diagonal blocks of $b$, when
we choose a suitable $\widetilde{\mathfrak{h}}$. 
We show examples of selections of $\widetilde{\mathfrak{h}}$ in
  Section~\ref{sec:detail}.  
\item{4.} Compute $m=pyp^\dag$. Here, $m \in \exp(\mathfrak{m})$
  because 
$\Theta(m)=
  \Theta(p)\Theta(y)\Theta(p^\dag)\: = \: py^\dag p^\dag \: = \:
  m^\dag$.
\item{5.} Compute $k=gm^\dag$. Then, $k$ always satisfies $k \in
  \exp(\mathfrak{k})$ because $(m^2)^\dag=g^\dag \Theta(g)$ and 
$\Theta(gm^\dag)=\Theta(g)\Theta(m^\dag)
= g(m^2)^\dag m
= gm^\dag$. 
\end{description}
Steps 2--4 provide a method for computing
the square root of a matrix to find $m$ from $m^2$.  
After these procedures, we obtain
\begin{eqnarray}
g & = & kpyp^\dag \: = \: \widetilde{k}yp^\dag. \label{msd}
\end{eqnarray}
Here, $\Theta({\widetilde{k}})=\widetilde{k}$, $\Theta(y)=y^\dag$, and
$\Theta(p^\dag)=p^\dag$, so that the decomposition follows the $G=KAK$
theorem. All matrices that appear through the algorithm can be
described using $g$ and $\Theta$. This enables us to show that
algorithm can automatically reproduce the well-known QFT circuit
(See, Section~\ref{sec:qft-msd}). 

\subsection{ Methods for performing steps 2 and 3 based on KGD
  \label{sec:detail}}  
Above, we provided a method for
computing $G=KAK$ decomposition. We did not provide a concrete
method for computing $p$, $b$, and $y$ in steps 2 and 3 in our
algorithm. To show examples of such concrete methods, we fixed $\Theta$ and
$\widetilde{\mathfrak{h}}$. Here, we treat examples that
compute KGD. We choose $\Theta$ as in (\ref{def-theta}) and show
two methods that compute $p$, $b$, $y$ for particular selections of
$\widetilde{\mathfrak{h}}$; i.e.,  
$\widetilde{\mathfrak{h}_1} = \mathrm{span} \; \{
|j \rangle \langle j| \otimes i\sigma_x 
\: | \: j=0,\cdots, 2^{n-1}-1 \}$ 
and 
$\mathfrak{\widetilde{h}_2}=\mathrm{span} \: \{ 
|j \rangle \langle j| \otimes 
   i(\sigma_{x} \otimes \sigma_{x}+ \sigma_{y} \otimes \sigma_{y})
\: | \: j=0,\cdots, 2^{n-2}-1 \}$.
We should note that the second selection is not always possible because
it demands that all eigenvalues appearing in $m^2$ in step 1
  should be duplicated twice. 
 
\subsubsection{Example 1  \label{sec:exp1}}
Let $\Theta$ be as in (\ref{def-theta}), and $\widetilde{\mathfrak{h}_1}$ as 
$\mathrm{span} \; \{|j \rangle \langle j| \otimes i\sigma_x 
\: | \: j=0,\cdots, 2^{n-1}-1 \}$. 
Then, we compute $p$, $b$, and $y$ in steps 2 and 3 as follows:  
\begin{description}
\item{(i)} Compute eigenvalue decomposition of $m^2$. \\
Let $m^2=\widetilde{p}d\widetilde{p}^\dag$ be eigenvalue decomposition
and $\mu_1, \cdots, \mu_{N}$ be the columns of $\widetilde{p}$, where
$N=2^n$.
Then, $d$ is a diagonal matrix in which diagonals have eigenvalues of
$m^2$ and all $\mu_j$'s are eigenvectors of $m^2$ and mutually orthogonal.  
\item{(ii)} Normalize all $\mu_j$'s by 
\begin{alignat}{2}
\nu_{2j-1}
&= \frac{\mu_j + \sigma_{nz}\mu_j}{\|\mu_j + \sigma_{nz}\mu_j\|},&\qquad
\nu_{2j}
&= \frac{\mu_j - \sigma_{nz}\mu_j}{\|\mu_j - \sigma_{nz}\mu_j\|}.
\end{alignat}
\item{(iii)} For all $\nu_j$'s that are associated with imaginary
  eigenvalues,  \\
\begin{description}
\item{(a)} let $W_1$, $W_2$, and $W_3$ be sets of vectors such that 
\begin{alignat*}{3}
W_1 &= \{\nu_j \mid \sigma_{nz}\nu_j = \nu_j\},&\quad
W_2 &= \{\nu_j \mid \sigma_{nz}\nu_j = -\nu_j\},&\quad
W_3 &= \{\nu_j \mid \sigma_{nz}\nu_j \neq \pm \nu_j\}.
\end{alignat*}
\item{(b)} For each $w \in W_3$, compute 
\begin{alignat}{2}
\nu^{+}
&= \frac{w + \sigma_{nz}w}{\|w + \sigma_{nz}w\|},&\qquad
\nu^{-}
&= \frac{w - \sigma_{nz}w}{\|w - \sigma_{nz}w\|}.
\end{alignat}
Here, $\| \cdot \|$ denotes the length of a vector. Then,  
\begin{itemize}
\item if all elements in $W_1$ and $u^{+}$ are linearly
  independent, then $W_1 = W_1 \: \cup \: \{ \nu^{+} \}$; 
\item if all elements in $W_2$ and $u^{-}$ are linearly
  independent, then $W_2 = W_2 \: \cup \: \{ \nu^{-} \}$.
\end{itemize}
\end{description}
\item{(iv)} Repeat steps (a) and (b) for all $\mu_j$'s
  that are associated with positive real eigenvalues. 
\item{(v)} Repeat steps (a) and (b) for all $\mu_j$'s
  that are associated with negative real eigenvalues.  
\item{(vi)} Let $p=(\upsilon_1, \upsilon_2, \cdots, \upsilon_{N})$,
  where $\upsilon_{2j-1} \in W_1$ and $\upsilon_{2j} \in W_2$, for
  $j=1,\cdots, N/2$.  \\ 
The computation procedure follows from Appendix~A. 
Since $\sigma_{nz}\upsilon_{2j-1}=\upsilon_{2j-1}$ and 
$\sigma_{nz}\upsilon_{2j}=- \upsilon_{2j}$,
we can easily check that $\Theta(p)=p$. 
Then, $b=p^\dag m^2 p$ satisfy $\Theta(b)=b^\dag$ and $b \in
\exp(\widetilde{\mathfrak{h}})$, where $b =
\sum_{j=0}^{2^{n-1}} |j \rangle \langle j| \otimes R_x(2\zeta_j)$ 
($0 \leq \zeta_j < \pi$). 
\item{(vii)} Compute $y$ by replacing all $R_x(2\zeta_j)$ appears in $b$
  with $R_x(\zeta_j)$. \\
Since $\Theta(R_x(2\zeta_j))=R_x^\dag(2\zeta_j)$,
  $R_x^2(\zeta_j)=R_x(2\zeta_j)$, then $y$ satisfies $y^2=b$ and $y
  \in \exp(\widetilde{\mathfrak{h}}_1)$.  
\end{description}
In step (vi), one may notice that, when $R_x(\pi)$ appears in $b$, then
we can use $R_y(\pi/2)$ instead of $R_x(\pi/2)$ as a replacement
rule. 

Since $\widetilde{k}$ and $p^\dag$ $\in$
are elements of $\exp(\mathfrak{k})$ that has the specific pattern as
shown in Fig.~\ref{pattern}, they have the following decomposition: 
\begin{alignat}{2}
\widetilde{k} &= g_1^{(0)} \otimes |0\rangle \langle 0| +  
       g_1^{(1)} \otimes |1\rangle \langle 1|, &\qquad
p^\dag &= g_2^{(0)} \otimes |0\rangle \langle 0| +  
       g_2^{(1)} \otimes |1\rangle \langle 1|, \label{decomp-k}
\end{alignat}
where $g_1^{(j)}, g_2^{(j)} \in SU(2^{n-1})$ for $\widetilde{k}, p^\dag
\in SU(2^n)$ ($j=0$ or $1$). 
Here, $g_1^{(0)}$ and $g_2^{(0)}$ are composed of nonzero elements
(black squares in Fig.~\ref{pattern}) of odd rows, and $g_1^{(1)}$ and
$g_2^{(1)}$ are composed of nonzero elements of even rows.  

\begin{figure}[tbp]
\vspace*{13pt}
\centerline{\epsfig{file=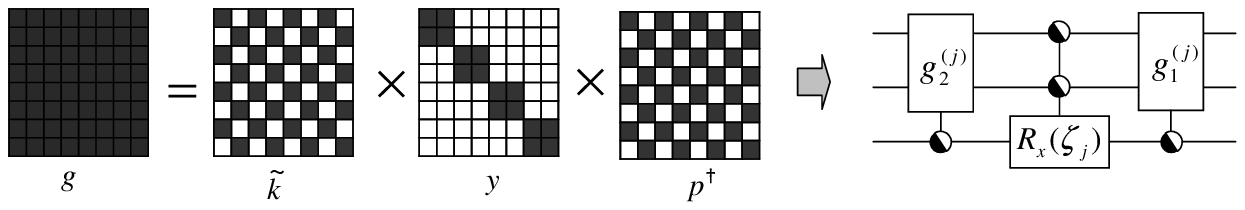}}
\vspace*{13pt}
\fcaption{Image of a decomposition when we choose $\Theta$ as 
in (\ref{def-theta}) and a Cartan subalgebra $\mathfrak{h}_1$ as 
$\mathrm{span} \; \{|j \rangle \langle j| \otimes i\sigma_x 
\: | \: j=0,\cdots, 2^{n-1}-1 \}$: \\
The matrices uses the same notation as in Fig.~\ref{pattern} to
represent the properties. In the quantum circuit, the symbol of the
control qubit represents the uniformly controlled
rotations~\cite{hut2,hut3}. $g_{\ell}^{(j-1)}\in SU(4)$ ($\ell, j=1$
or $2$) are applied selectively; that is, $g_{1}^{(0)}$ and $g_{2}^{(0)}$
are applied when the third qubit is $|0\rangle$, whereas
$g_{1}^{(1)}$ and $g_{2}^{(1)}$ are applied when it is $|1\rangle$.
\label{msd-image}} 
\end{figure}

Fig.~\ref{msd-image} shows the image of a decomposition in
(\ref{kak}) for a three-qubit system; that is, we choose
$\Theta$ as in (\ref{def-theta}) and a Cartan subalgebra as
$\mathfrak{h}_1$. 
Applying the decomposition in (\ref{msd}) recursively to elements
$g_1^{(j)}, g_2^{(j)} \in SU(2^{n-1})$ ($j$=$0$ or $1$), we obtain a
sequence of uniformly controlled rotations like in Fig.~13 in
Ref.~\cite{hut2}, except that $R_x$ is used instead of $R_y$
in our case. The full decomposition of these uniformly controlled
rotations into elementary gates has been 
provided by M\"{o}tt\"{o}nen~{\it et al.}~\cite{hut2, hut3}.
Also, if we change the order of qubits and apply the quantum
Multiplexor decomposition to $\widetilde{k}$ and $p^\dag$ in
(\ref{kak}), the produced circuit is the same as that in Fig.~2 in
Ref.~\cite{shende2}, except that rotation $R_y$ is used instead of
$R_x$. 
Therefore, the number of elementary gates needed to compose $g\in
SU(2^n)$ in our method is $O(4^n)$, which is the same as in
Refs.~\cite{hut1, shende2}.

\subsubsection{Example 2 \label{sec:exp2}}
Here, we show another example of methods for computing $p$, $b$, and $y$
in Section~\ref{sec:alg}. Here, we choose $\mathfrak{\widetilde{h}_2}=
\mathrm{span} \: \{ |j \rangle \langle j| \otimes
    i(\sigma_{x} \otimes \sigma_{x} + \sigma_{y} \otimes \sigma_{y})
\: | \: j=0,\cdots, 2^{n-1}-1 \}$. 
The Cartan involution $\Theta$ is the same as in (\ref{def-theta}). 
The decomposition of this type is chosen as an example of a
decomposition of the QFT.  
In this case, we only change steps (vi) and (vii) in
Section~\ref{sec:exp1} as follows: 
\begin{description}
\item{(vi)} Let $p=(\upsilon_1, \upsilon_2, \cdots, \upsilon_{N})$, 
for $j=1,\cdots, N/4$,      
\begin{itemize}
\item $\nu_{4j-3} \in W_1$ and it is associated with positive
  eigenvalues,
\item $\nu_{4j-2} \in W_2$ and it is associated with negative
  eigenvalues, 
\item $\nu_{4j-1} \in W_1$ and it is associated with negative
  eigenvalues, 
\item $\nu_{4j} \in W_2$ and it is associated with positive eigenvalues.
\end{itemize}
Then, $p$ also satisfies $\Theta(p)=p$, and $b=p^\dag m^2p$ is a
block-diagonal matrix, in which the diagonals are constructed from
the $4\times 4$ matrix  
\begin{eqnarray*}
\begin{pmatrix}
1 & 0 & 0 & 0 \cr
0 & \cos(2\zeta_j) & i \sin(2 \zeta_j) & 0 \cr
0 & i \sin(2 \zeta_j) & \cos(2\zeta_j) & 0 \cr
0 & 0 & 0 & 1
\end{pmatrix} = \exp(i \zeta_j (\sigma_x\otimes \sigma_x + \sigma_y
\otimes \sigma_y)).
\end{eqnarray*}
Here, the middle part of the above matrix is $R_x(2\zeta_j)$. 
\item{(vii)} Compute $y$ by replacing $R_x(2\zeta_j)$ in $b$ with
  $R_x(\zeta_j)$.  
\end{description}

\section{Decompositions of the QFT using $G=KAK$ decomposition
  \label{sec:QFT}} 
In this section, we show that we can automatically reproduce the well
known QFT circuit using our method.  
All the matrices that appear through our algorithm can be
described using the input matrix $g$ and $\Theta$. In contrast, it is
difficult to describe all the matrices that appear through the
CSD-based algorithm because, as shown in (\ref{def-csd}), 
the input matrix $g$ has to be divided into four square matrices and
SVD has to be applied to each partitioned matrix. Furthermore, 
we have to choose a suitable decomposition for each partitioned matrix
$g_{11}$, $g_{12}$, $g_{21}$, and $g_{22}$ such that
Eq.~(\ref{def-csd}) holds, which makes it difficult to formulate
$U$, $\Sigma$, and $V$ in (\ref{def-csd}).

Fortunately, the $n$-qubit QFT is a very special matrix that has the
following property: If we permute the order of qubits, then each
partitioned matrix can be described using ($n-1$)-qubit QFT. Using the
feature, we provide a decomposition of the QFT by CSD-based
algorithm. This is shown in Section~\ref{sec:qft-csd}. 

\subsection{Notation \label{sec:standard-qft}} 
\noindent
The QFT on $n$ qubits, $F_{n}$, is a $2^n \times 2^n$ matrix such that  
\begin{eqnarray}
F_{n}& = & \left(\frac{1}{\sqrt{2^n}}\omega^{(j-1)(\ell-1)}_{n}
\right)_{j\ell},  
\qquad \mathrm{where} \; \; \omega_{n}=\exp \left( \frac{2\pi i}{2^n}
\right).
\label{qft-def}
\end{eqnarray}
We define $Q_n$ as a $2^n \times 2^n$ permutation matrix: $Q_n\: = \:
\chi_{n-1}^{n} \cdots \chi_2^{n} \chi_1^{n}$, where $\chi_j^k$ is the
SWAP gate applied to the  $j$-th and the $k$-th qubits. Let
$H_1=H\otimes I^{\otimes (n-1)}$, then (\ref{qft-def})
is written as  
\begin{eqnarray}
F_{n}&=&\frac{1}{\sqrt 2}
\begin{pmatrix}
F_{n-1}&\Omega_{n-1}F_{n-1}\\
F_{n-1}&-\Omega_{n-1}F_{n-1}
\end{pmatrix}Q_n
\; =\;  H_1 D_n (I \otimes F_{n-1}) Q_n, \label{qft-decomp}
\end{eqnarray}
where
\begin{alignat*}{2}
D_n &= \begin{pmatrix}
I^{\otimes (n-1)} &0\\
0&\Omega_{n-1}
\end{pmatrix}, &\qquad 
\Omega_{n-1}&= diag \: ( 1, \omega_{n}, \cdots, \omega_n^{2^{n-1}-1}).
\end{alignat*}
This notation follows from Section~4.6.4 in Ref.~\cite{golub}. 

\subsection{Decomposition of the QFT by our method
  \label{sec:qft-msd}} 
\begin{figure}[tbp]
\vspace*{10pt}
\centerline{\epsfig{file=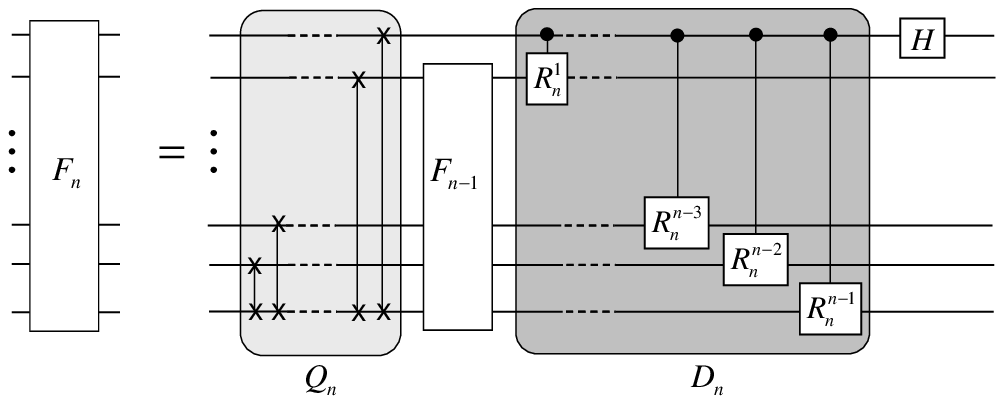}}
\vspace*{5pt}
\fcaption{Decomposition of the QFT by our algorithm. 
Here, $R_n^j = diag \:(1, \omega_n^j)$, i.e., a single-qubit rotation,
and $Q_n$ is composed of $n-1$ SWAP gates.  \label{qft-circuit}} 
\end{figure}

Following Section~\ref{sec:alg}, we compute a decomposition as follows: 
\begin{description}
\item{1.} Compute $m^2=\Theta(F_n^\dag)F_n$. \\
Let $S$ be $(I \otimes F_{n-1}) Q_n$. 
Since $\Theta(H_1)=H_1$, $\Theta(D_n)=D_n$,
and $\sigma_{1z}S = S 
\sigma_{nz}$, we have $m^2=S^\dag \sigma_{1z} \sigma_{nz} S$.   
All column vectors of $S^\dag$ are then eigenvectors of $m^2$ because
$\sigma_{1z} \sigma_{nz}$ is a diagonal matrix in which diagonal
elements are eigenvalues of $m^2$.  
\item{2.} Decompose $m^2=pbp^\dag$ such that $\Theta(p) =p$ and
  $\Theta(b)=b^\dag$. \\
We define $p=S^\dag Q_n$ (This selection was done so that $p$ satisfies
  $\Theta(p)=p$ and follows Section~\ref{sec:exp2}), then $b \:= \:
  p^\dag m^2 p \: = \: 
Q_n^\dag\sigma_{1z}\sigma_{nz}Q_n \: = \: I^{\otimes (n-2)}
\otimes diag ( 1, \: -1, \: -1, \: 1)$.
\item{3.} Choose $y$ such that $\Theta(y) = y^\dag$. Following the 
step (vi) in Section~\ref{sec:exp2}, we have \\ 
\begin{eqnarray*}
y & = &  I^{\otimes (n-2)} \otimes
\begin{pmatrix}
1&0&0&0\\
0&0&i&0\\
0&i&0&0\\
0&0&0&1
\end{pmatrix} 
\: = \: I^{\otimes (n-2)} \otimes
\exp (\pi (\sigma_x 
\otimes \sigma_x + \sigma_y \otimes \sigma_y)/4). 
\end{eqnarray*}
This is obtained by replacing each $4\times 4$ diagonal block of $v$,
$b=diag ( 1, \: -1, \: -1, \: 1)$, with $y=\exp (\pi (\sigma_x 
\otimes \sigma_x + \sigma_y \otimes \sigma_y)/4)$. 
We can easily check that $\Theta(y)=y^\dag$ and $y^2=b$, because
$\Theta(y)=y^\dag$ and $y^2=b$. 
\item{4.} Compute $m\: = \: pyp^\dag$. 
\item{5.} Compute $k \: = \: gm^\dag \: = \: F_n m^\dag$. 
\end{description}
Then, $\widetilde{k}\: = \: kp \: = \: (F_{n}py^\dag
p^\dag)p=F_{n}py^\dag$, so we have the following decomposition: 
\begin{eqnarray}
F_n & = & \widetilde{k}yp^\dag \: = \: (H_1 D_n Q_n y^\dag) y
(Q_n^\dag S) \: = \: H_1 D_n S \: = H_1 D_n (I \otimes F_{n-1})Q_n.
\label{decomp-by-msd} 
\end{eqnarray} 
We apply a similar decomposition to $F_{j}$, for $j=n-1,
n-2,\cdots, 2$. Next, we show a decomposition of $D_n$. 
$D_n$ is controlled-$\Omega_{j}$ (where $j=n-1, n-2, \cdots, 2$), so
it suffices to consider the decomposition of $\Omega_{j}$. 
Since $\Omega_j \in \exp(\mathfrak{k})$ (it follows
from $\Theta(\Omega_j)=\Omega_j$), we apply the decomposition in
(\ref{decomp-k}) to $\Omega_j$. 
Consider $\Omega_3$ as an example,
then we have $\Omega_3 = g_1^{(0)}\otimes |0 \rangle \langle 0| 
+ g_1^{(1)} \otimes|1 \rangle \langle 1|$, where 
$g_1^{(0)} = diag \: (1, \omega^2, \omega^4, \omega^6)$ and 
$g_1^{(1)} = diag \: ( \omega, \omega^3, \omega^5, \omega^7)= 
\omega \cdot diag \: (1, \omega^2, \omega^4, \omega^6)$. 
Then, we have $
 \Omega_3=  diag \: ( 1, \omega^2, \omega^4, \omega^6) \otimes 
diag \: (1, \omega)$. 
Similarly, since $diag \:( 1, \omega^2, \omega^4, \omega^6)$ is also
an element of $\exp(\mathfrak{k})$, it is decomposed into 
$diag\: (1,\omega^4) \otimes  diag \: (1, \omega^2)$.
Therefore, $\Omega_{n-1}$ is composed of $n-1$ single-qubit rotations
as follows:   
\begin{eqnarray}
\Omega_{n-1} & = & 
\begin{pmatrix}
1 & 0 \\
0 & \omega^{2^{n-1}-1}_{n}
\end{pmatrix} 
\otimes  \cdots \otimes 
\begin{pmatrix}
1 & 0 \\
0 & \omega^{2^{j-1}}_{n}
\end{pmatrix} \otimes \cdots \otimes
\begin{pmatrix}
1 & 0 \\
0 & \omega_{n}^2
\end{pmatrix} \otimes 
\begin{pmatrix}
1 & 0 \\
0 & \omega_{n}
\end{pmatrix}.  \label{decomp-Omega}
\end{eqnarray}
The circuit obtained from the above decomposition is shown in
Fig.~\ref{qft-circuit}. We apply a similar decomposition to $F_{j}$ 
(for $j=n-1,n-2,\cdots,2$). 
Finally, we have a full QFT decomposition composed of $n$ Hadamard
gates, $\frac{1}{2}n(n-1)$ controlled-rotations, and 
$\lfloor \frac{n}{2} \rfloor$ SWAP gates. 
Here, SWAP gates that appeared in a sequence of permutations $Q_n Q_{n-1}
\cdots Q_2$ were optimized.
It is known that a controlled-rotations can be implemented by
three single-qubit rotation and two CNOTs, so that the number of elementary 
gates that appear in Fig.~\ref{qft-circuit} is $O(n^2)$.  
 
\subsection{Decomposition of the QFT by CSD \label{sec:qft-csd}}
If we apply $Q_n^\dag$ to the input matrix $F_n$ beforehand,
then we have the following decomposition:  
\begin{eqnarray}
F_n Q_n^\dag & = & U \Sigma V \nonumber \\
& = & 
\begin{pmatrix}
I^{\otimes {n-1}} & 0 \\
0 & I^{\otimes {n-1}} 
\end{pmatrix} \cdot \frac{1}{\sqrt{2}} 
\begin{pmatrix}
I^{\otimes (n-1)} & -I^{\otimes (n-1)} \\
I^{\otimes (n-1)} & I^{\otimes (n-1)} 
\end{pmatrix} \cdot 
\begin{pmatrix}
F_{n-1} & 0 \\
0 & -\Omega_{n-1}F_{n-1} 
\end{pmatrix}, \label{decomp-by-csd} \\
& = & (H\sigma_z \otimes I^{\otimes (n-1)})
(\sigma_z\otimes
I^{\otimes (n-1)}) D_n (I \otimes F_{n-1}) \: = \:
H_1 S.
\end{eqnarray}
We can easily check that (\ref{decomp-by-csd}) satisfies the 
definition of CSD, where all $\theta_j$'s appear in $c$ and $s$ in
(\ref{def-csd}) are $\pi/4$. Although we have to apply $Q_{j}^\dag$
beforehand when we apply CSD recursively to $F_{j}$
($j=1,\cdots,n-1$), we obtain the well-known QFT circuit. 
Using the feature, Tucci reproduced the well-known QFT circuit using a
CSD-based algorithm~\cite{tucci1, tucci2}. 

\subsection{Decomposition of the QFT by the quantum Shannon
  decomposition (QSD) \label{sec:qft-qsd}}
\begin{figure}[tbp]
\vspace*{10pt}
\centerline{\epsfig{file=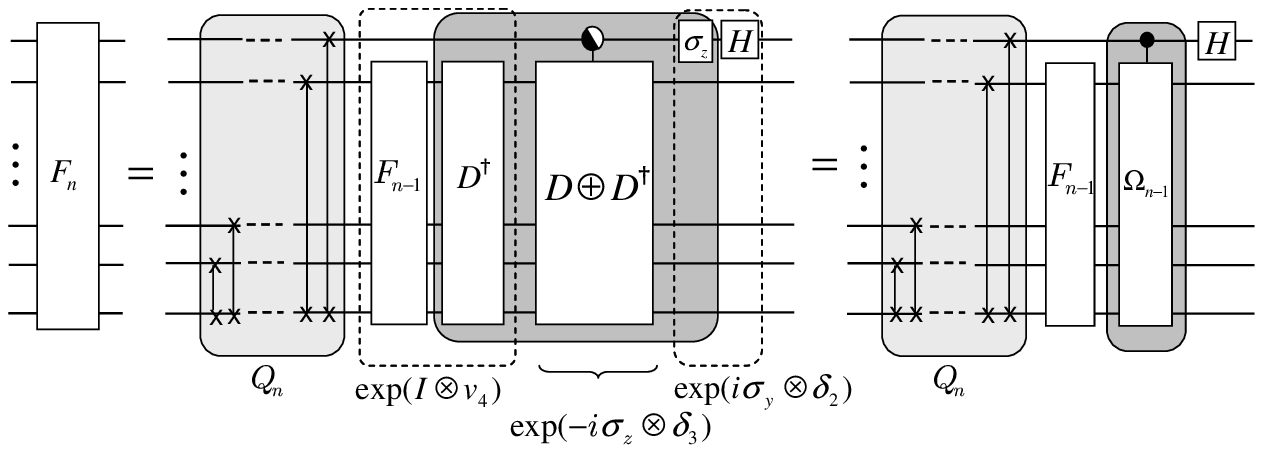}}
\vspace*{5pt}
\fcaption{Decomposition of the QFT by the QSD. 
Here, we move $Q_n^\dag$ to the right-hand-side by inverting it. 
The dark-gray block in the middle circuit can be simplified.
We show the simplified block by the same color in the rightmost circuit. 
Here, $D=\sqrt{-\Omega_{n-1}}$.  \label{qsd-circuit}} 
\end{figure}

The QSD~\cite{shende2} is a method that combines CSD and the quantum
multiplexor decomposition. First, we compute CSD for the input
matrix and then apply the quantum multiplexor
decomposition (cf.~\cite{shende2}, Theorem~12), 
\begin{eqnarray}
\begin{pmatrix}
u_1 & 0 \\
0 & u_2 
\end{pmatrix} & = & 
\begin{pmatrix}
v & 0 \\
0 & v 
\end{pmatrix} 
\begin{pmatrix}
d & 0 \\
0 & d^\dag
\end{pmatrix} 
\begin{pmatrix}
w & 0 \\
0 & w 
\end{pmatrix}, \label{q-multiplexor}
\end{eqnarray}
to $U$ and $V$ in (\ref{def-csd}). Here,
 $u_1u_2^\dag=vd^2v^\dag$ and $w=dv^\dag u_2$. 
In the QFT, we apply the decomposition in (\ref{q-multiplexor}) to
 $V$ because $U$ is an identity matrix in (\ref{decomp-by-csd}). 
Then, we have $v=I^{\otimes (n-1)}$ and
 $d=\sqrt{-\Omega_{n-1}^\dag}$, and $w=\sqrt{-\Omega_{n-1}}F_{n-1}$ in
 (\ref{q-multiplexor}). Here,  
$\sqrt{-\Omega_{n-1}}$ is a $2^{n-1}\times 2^{n-1}$ diagonal matrix
 whose $(j,j)$-th component is $i \omega_{n+1}^{j-1}$, ($j=1,\cdots,
2^{n-1}$). Therefore, the decomposition of QFT by the QSD is as
 follows: 
\begin{eqnarray}
F_n Q_n^\dag & = & \exp(i \sigma_y \otimes \delta_2) \exp(-i \sigma_z
\otimes \delta_3) (I \otimes v_4),
\label{decomp-by-qsd}
\end{eqnarray}
where $v_4 = \sqrt{-\Omega_{n-1}}F_{n-1}$ and $\delta_2$ and
$\delta_3$ are $2^{n-1} \times 2^{n-1}$ diagonal matrices.
Each element of $\delta_2$ is $\pi/2^{n-1}$ and each $(j,j)$-th
element of $\delta_3$ is $-j\pi/2^{n+1}$. 
Furthermore, $\sqrt{\Omega_{n-1}}$ is composed of $n-1$ single-qubit
rotations as follows:
\begin{eqnarray*}
\sqrt{-\Omega_{n-1}} & = & i
\begin{pmatrix}
1 & 0 \\
0 & \omega_{n+1}^{2^{n-1}}
\end{pmatrix} 
\otimes  \cdots \otimes 
\begin{pmatrix}
1 & 0 \\
0 & \omega_{n+1}^{2^{j-1}}
\end{pmatrix} \otimes \cdots \otimes
\begin{pmatrix}
1 & 0 \\
0 & \omega_{n+1}^2
\end{pmatrix} \otimes 
\begin{pmatrix}
1 & 0 \\
0 & \omega_{n+1}
\end{pmatrix}.  
\end{eqnarray*}
Here, (\ref{decomp-by-qsd}) is also equal to the well-known QFT
decomposition (\ref{qft-decomp}) after optimization as shown in
Fig.~\ref{qsd-circuit}. 
To obtain the decomposition in (\ref{decomp-by-qsd}), note that we
have to apply $Q_j$ ($j=0,\cdots, n$) beforehand for $F_j$. 

\section{Conclusion}\label{sec:conc}
\noindent 
We introduced a new algorithm for computing any type-{\bf AIII} $KAK$
  decomposition according to the given global Cartan involution $\Theta$. 
Recursively performing the decomposition leads us to a quantum circuit
  composed of uniformly controlled rotations. 
The algorithm can derive any matrix decomposition corresponding to the
  type-${\bf AIII}$ $KAK$ decomposition, and it contains CSD and KGD as
  its special cases. 
This is because our algorithm contains arbitrariness in selecting the
  Cartan subalgebra $\mathfrak{h}$ and a square root matrix $m$ for
  the given Cartan involution $\Theta$, where $m$ is a matrix derived 
from a global Cartan decomposition $g=km$. 
We also showed two methods for computing a square root
  matrix. 

Although the correctness of our algorithm depends on Lie group theory,
the main methods involved are eigenvalue decomposition
and a simple replacement rule. Thus, we can compute a decomposition 
without knowledge of Lie group theory.

As an example, we showed that our method automatically reproduces 
the well-known QFT circuit for arbitrary $n$-qubits. 
When using CSD-based algorithms, we have to apply some permutations 
beforehand in order to reproduce
the circuit. The same technique can not always be used to describe the canonical form of the decomposition for a given matrix because matrices do not always have a convenient form like QFT.
Our algorithm might be useful in showing the effectiveness of
$G=KAK$ matrix decompositions for other particular input matrices,
because all matrices appearing through our algorithm can be described
using a given input matrix $g$ and a given Cartan involution $\Theta$.   

\nonumsection{References}

\appendix{~~Properties of eigenvalues of $m^2=\Theta(G^\dag)G$}
In Section~\ref{sec:exp1}, we show that $b$ is a block-diagonal matrix,
each block of which is $R_x(2\zeta)$, where $\widetilde{\mathfrak{h}}=
\mathrm{span} \{ |j\rangle \langle j| \otimes i \sigma_x 
\: | \: j=0,\cdots, 2^{n-1}-1 \} \subset \mathfrak{m}$. 
This appendix provide two lemmas and proofs for the eigenvalues of
$m^2=\Theta(G^\dag)G$, where $G\in SU(2^n)$ to prove that.  
Let $X=m^2=\Theta(G^\dag)G$. 

\begin{lemma}[Properties of complex eigenvalues of
    $X$]\label{lem:comp_ev}

\mbox{}
\begin{enumerate}
\item
The number of the complex eigenvalues of $X$, with multiplicity
counted, is even.
They are of the form $\alpha_1$, $\overline{\alpha_1}$, 
$\alpha_2$, $\overline{\alpha_2}$, \dots, $\alpha_t$, $\overline{\alpha_t}$,
repeated with multiplicity.

\item
There exists unit length vectors $u_1$, \dots, $u_{2t}$ that are
mutually orthogonal and satisfy $\sigma_{nz}u_{2j-1} = u_{2j-1}$ and 
$\sigma_{nz}u_{2j} = -u_{2j}$.
Furthermore, there exists $\zeta_1$, \dots, $\zeta_t\in {\mathbb R}$
satisfying $X u_{2j-1} = \cos(\zeta_j)u_{2j-1} -
i\sin(\zeta_j)u_{2j}$ and $X u_{2j} = -i\sin(\zeta_j)u_{2j-1} +
\cos(\zeta_j)u_{2j}$. 
\end{enumerate}
\end{lemma} \vspace{9pt}
\begin{proof}
Let $\alpha_1$, \dots, $\alpha_{2^n}$ be the eigenvalues of $X$,
repeated with multiplicity.
Then, except for the order, the elements of the sequence
$\overline{\alpha_1}$, \dots, $\overline{\alpha_{2^n}}$ are equal to  
$\alpha_1$, \dots, $\alpha_{2^n}$ because $\Theta(X)=X^\dagger$.
Therefore, the number of the complex eigenvalues, counted with
multiplicity, is even, and we denote it as $2t$.
We can write the complex eigenvalues as $\alpha_1,\
\overline{\alpha_1},\ \dots, \alpha_t,\ \overline{\alpha_t}$. 
That is, the first statement holds.

Let $u$ be an eigenvector of $X$ corresponding to the eigenvalue
$\alpha$, then $\sigma_{nz}u$ is an eigenvector of $X$ corresponding
to the eigenvalue $\overline{\alpha}$ because $X(\sigma_{nz}u) =
\sigma_{nz}X^\dagger u = \overline{\alpha_j}(\sigma_{nz}u)$.
Now, let $W$ be the eigenspace corresponding to a complex eigenvalue
$\alpha$ and $\beta_1$, \dots, $\beta_{r}$ be an orthonormal basis of
$W$. Then, let $\sigma_{nz}\beta_1$, \dots, $\sigma_{nz}\beta_r$ be an
orthonormal basis of the eigenspace corresponding to the eigenvalue
$\overline{\alpha}$.
Therefore, $u_{2j-1} = \beta_{j} +\sigma_{nz}\beta_j$ and
$u_{2j}= \beta_{j} - \sigma_{nz}\beta_{j}$ for $j=1,\cdots, r$, span
the eigenspace of $X$ corresponding to the eigenvalues $\alpha$ and
$\overline{\alpha}$. 
Furthermore, they are eigenvectors of $\sigma_{nz}$ because
$\sigma_{nz}u_{2j-1} = u_{2j-1}$ and $\sigma_{nz}u_{2j} = -u_{2j}$.
On the other hand, the following relations hold: 
\begin{eqnarray*}
Xu_{2j-1} &= &\alpha \beta_j + \overline{\alpha}\sigma_{nz} \beta_j 
\: = \: \frac{\alpha + \overline{\alpha}}{2}u_{2j-1}
	+ \frac{\alpha - \overline{\alpha}}{2}u_{2j}
\: = \: \RE(\alpha)u_{2j-1} + i\IM(\alpha)u_{2j}, \\
Xu_{2j} &= &\alpha \beta_j - \overline{\alpha}\sigma_{nz} \beta_j 
\: = \: \frac{\alpha - \overline{\alpha}}{2}u_{2j-1}
	+ \frac{\alpha + \overline{\alpha}}{2}u_{2j} 
\: = \: i\IM(\alpha)u_{2j-1} + \RE(\alpha)u_{2j}.
\end{eqnarray*}
Therefore, put $\zeta=-\arg(\alpha)$. Then
$\cos(\zeta)=\RE(\alpha)$ and $\sin(\zeta)=-\IM(\alpha)$.
That is,
\[
X u_{2j-1} = \cos(\zeta)u_{2j-1} - i\sin(\zeta)u_{2j},\qquad
X u_{2j} = -i\sin(\zeta)u_{2j-1} + \cos(\zeta)u_{2j}.
\]
Similar arguments hold for the other complex eigenvalues.
\end{proof} \vspace{9pt}

\begin{lemma}[Properties of real eigenvalues of $X$]

\label{lem:real_ev}
Real eigenvalues of $X$ are $\pm 1$.
\begin{enumerate}
\item
The multiplicity of the eigenvalue $1$ is even.
There exists an orthonormal basis $u_{2t+j}$ ($j=1, \cdots,2\mu$) of
the eigenspace corresponding to the eigenvalue $1$ that satisfies the
$\sigma_{nz}u_{2t+2j-1} = u_{2t+2j-1}$ and $\sigma_{nz}u_{2t+2j} =
-u_{2t+2j}$. 

\item
The multiplicity of the eigenvalue $-1$ is even.
There exists an orthonormal basis $u_{2t+2\mu+j}$ ($j=1,\cdots,2\nu$)
of the eigenspace corresponding to the eigenvalue $-1$ that satisfies \\
$\sigma_{nz}u_{2t+2\mu+2j-1} = u_{2t+2\mu+2j-1}$ and
$\sigma_{nz}u_{2t+2\mu+2j} = -u_{2t+2\mu+2j}$.
\end{enumerate}
\end{lemma} \vspace{9pt}
\begin{proof}
Lemma~\ref{lem:comp_ev} implies that the product of all the complex 
eigenvalues is $1$, and thus the product of all the real eigenvalues
is $1$. Therefore, the real eigenvalues of $X$ are $1$ or $-1$, thus both
of the multiplicities of the eigenvalues $1$ and $-1$ are even.
Let $W_1$ and $W_2$ be the eigenspaces of $X$ corresponding to the
eigenvalues of $1$ and $-1$, respectively.
Then $\sigma_{nz}W_1\subset W_1$ and $\sigma_{nz}W_2\subset W_2$ hold
because $\Theta(X)=X^\dagger$.
Put $W=W_1\oplus W_2$. Then $W^\bot$is the direct sum of the
eigenspaces for the complex eigenvalues of $X$.
The trace of $\sigma_{nz}|_{W}$ is $0$ because Lemma~\ref{lem:comp_ev}
implies that the trace of $\sigma_{nz}|_{W^\bot}$ is $0$, 
and this implies that the multiplicities of the eigenvalues of
$1$ and $-1$ of $\sigma_{nz}|_W$ are equal.
We write $W_1=W_{11}\oplus W_{12}$ and $W_2=W_{21}\oplus W_{22}$,
where $W_{11}$ and $W_{21}$ are the eigenspaces of $\sigma_{nz}$
corresponding to the eigenvalue $1$, and 
$W_{12}$ and $W_{22}$ are the eigenspaces of $\sigma_{nz}$
corresponding to the eigenvalue $-1$.
Let $\dim W_{ij}$ be $d_{ij}$. Then we have $d_{11}+d_{21} =
d_{12}+d_{22}$. 

We can make similar arguments for$X' = \Theta(G)G^\dag$.
We write
$W'_1=W'_{11}\oplus W'_{12}$ and $W'_2=W'_{21}\oplus W'_{22}$,
where $W'_{11}$ and $W'_{21}$ are the eigenspaces of $\sigma_{nz}$
corresponding to the eigenvalue $1$, and 
$W'_{12}$ and $W'_{22}$ are the eigenspaces of $\sigma_{nz}$
corresponding to the eigenvalue $-1$.
Let $\dim W'_{ij}$ be $d'_{ij}$. Then we have $d'_{11}+d'_{21}
=d'_{12}+d'_{22}$. 

From Lemma~\ref{lem:xx'} below, we have $d_{11} = d'_{11}$, $d_{12} =
d'_{12}$, $d_{21} = d'_{22}$, and $d_{22} = d'_{21}$. 
Then, we have $d_{11}=d_{12}$ and $d_{21}=d_{22}$. Therefore, the
statements of the lemma hold. 
\end{proof} \vspace{9pt}
%
\begin{lemma}{\label{lem:xx'}}
Let $X=\Theta(G^\dag)G$ and $X' = \Theta(G)G^\dag$.
\begin{enumerate}
\item
If $Xu=u$ and $\sigma_{nz}u=u$,
then $X'(Gu) = Gu$ and $\sigma_{nz}(Gu) = Gu$.

\item
If $Xu=u$ and $\sigma_{nz}u=-u$,
then $X'(Gu) = Gu$ and $\sigma_{nz}(Gu) = -Gu$.

\item
If $Xu=-u$ and $\sigma_{nz}u=u$,
then $X'(Gu) = -Gu$ and $\sigma_{nz}(Gu) = -Gu$.

\item
If $Xu=-u$ and $\sigma_{nz}u=-u$,
then $X'(Gu) = -Gu$ and $\sigma_{nz}(Gu) = Gu$.
\end{enumerate}
\end{lemma} \vspace{9pt}
\begin{proof}
First, we prove the statements for the eigenvalues and eigenvectors of
$X'$.
Since $G^\dagger X'G=\Theta(X)$ holds,
we have $(X'(Gu), Gu) = (G^\dagger X' Gu, u)
= (\Theta(X) u, u)
= (X^\dagger u, u)
= (u, Xu)$. 
Thus, the equation $X u = \epsilon u$, where $\epsilon=\pm 1$,
implies $(X'(Gu), Gu) = \epsilon(u, u)$.
On the other hand, the Cauchy-Schwarz inequality implies
$|(X'(Gu), Gu)| \leq \|X'(Gu)\|\cdot\|Gu\|$.
Since the right-hand side is equal to 
$\|Gu\|^2 =\|u\|^2 = (u, u)=|(X'(Gu), Gu)|$, the equality $|(X'(Gu),
Gu)|=\|X'(Gu)\|\cdot\|Gu\|$ holds.
Therefore,
we have $X'(Gu)= \alpha (Gu)$ for some $\alpha\in {\mathbb C}$;
that is, $Gu$ is an eigenvector of $X'$ corresponding to the
eigenvalue $\alpha$.
The equality $\alpha=\epsilon$ follows $\epsilon(u, u)=
(X'(Gu), Gu) = (\alpha Gu, Gu) = \alpha (Gu, Gu)
= \alpha (u,u)$.

To prove the statements for the eigenvalues and eigenvectors of
$\sigma_{nz}$, we use the following equations:
\[
(\sigma_{nz}(Gu), Gu) = (G^\dagger \sigma_{nz}Gu, u)
= (\sigma_{nz}X u, u)
= (X u, \sigma_{nz}u). 
\]
The equations $X u = \epsilon u$ and $\sigma_{nz}u = \epsilon' u$,
where $\epsilon$, $\epsilon'=\pm 1$, imply $(\sigma_{nz}(Gu), Gu) =
\epsilon\epsilon'(u, u)$. 
From similar arguments for the eigenvalues and eigenvectors of $X'$,
we have $\sigma_{nz}(Gu)= \epsilon\epsilon'(Gu)$.
\end{proof} 

\end{document}